\documentclass[conference]{IEEEtran}
\usepackage{epsfig,cite,bm}
\def\defequal{\stackrel{\mathrm{def}}{=}}
\def\argmax{\mathop{\mbox{arg\,max}}}
\IEEEoverridecommandlockouts

\begin{document}
\title{Opportunistic Scheduling and Beamforming for MIMO-OFDMA Downlink Systems with Reduced Feedback}
\author{Man-On Pun, Kyeong Jin Kim and H. Vincent Poor
\thanks{Man-On Pun and H. Vincent Poor are with the Department of Electrical Engineering, Princeton University, Princeton, NJ 08544.}
\thanks{Kyeong Jin Kim is with Nokia Inc, 6000 Connection Dr., Irving TX 75039.}
\thanks{This research was supported in part by the Croucher Foundation under a post-doctoral fellowship, and in part by the U. S. National Science Foundation under Grants ANI-03-38807 and CNS-06-25637.}}
\maketitle
\thispagestyle{empty}
\begin{abstract}
Opportunistic scheduling and beamforming schemes with reduced feedback are proposed for MIMO-OFDMA downlink systems.  Unlike the conventional beamforming schemes in which beamforming is implemented solely by the base station (BS) in a per-subcarrier fashion, the proposed schemes take advantages of a novel channel decomposition technique to perform beamforming jointly by the BS and the mobile terminal (MT). The resulting beamforming schemes allow the BS to employ only {\em one} beamforming matrix (BFM) to form beams for {\em all} subcarriers while each MT completes the beamforming task for each subcarrier locally. Consequently, for a MIMO-OFDMA system with $Q$ subcarriers, the proposed opportunistic scheduling and beamforming schemes require only one BFM index and $Q$ supportable throughputs to be returned from each MT to the BS, in contrast to $Q$ BFM indices and $Q$ supportable throughputs required by the conventional schemes. The advantage of the proposed schemes becomes more evident when a further feedback reduction is achieved by grouping adjacent subcarriers into exclusive clusters and returning only cluster information from each MT. Theoretical analysis and computer simulation confirm the effectiveness of the proposed reduced-feedback schemes.
\end{abstract}

\section{Introduction}\label{sec:intro}
Orthogonal frequency-division multiple-access (OFDMA) has recently attracted much attention as a promising technique for future broadband wireless communications. In an OFDMA downlink system, the base station (BS) transmits data to several active mobile terminals (MTs) simultaneously by modulating each MT's data onto an exclusive set of orthogonal subcarriers. In addition to its robustness to multipath fading and high spectral efficiency, OFDMA is particularly attractive due to its flexibility in allocating subcarriers to different MTs based on their different quality of service (QoS) requirements and channel conditions in a dynamic fashion \cite{Wong99}. Furthermore, the recent advancement in multiple-input multiple-output (MIMO) techniques has inspired considerable research interest in MIMO-OFDMA. However, perfect channel state information (CSI) is usually required at the BS in order to fully harvest the advantages provided by MIMO in the downlink transmission, which incurs formidable feedback overhead from MTs. This problem becomes particularly challenging in MIMO-OFDMA since the required feedback amount is proportional to the number of subcarriers. To circumvent this obstacle, an opportunistic scheduling and beamforming scheme has been proposed for single-carrier (SC) multiple-input single-output (MISO) systems in \cite{Tse02} as an effective means of achieving the asymptotic sum-rate capacity by exploiting {\em multiuser diversity} with limited channel feedback. Some extensions of \cite{Tse02} have been developed for SC-MIMO systems \cite{Kim05,Chung03}. In particular, \cite{Chung03} has proposed a singular value decomposition (SVD)-based scheme to schedule data transmission to the MT whose channel matrix has the right singular vectors aligning with beams in the common codebook shared by the BS and MTs. Most recently, an opportunistic scheme has been proposed for single-input single-output (SISO)-OFDMA by grouping the adjacent subcarriers into exclusive clusters \cite{Svedman04}. Assuming the subcarriers in each cluster have approximately the same channel conditions, \cite{Svedman04} has demonstrated good throughput performance by only feeding the average cluster SNRs from each MT back to the BS.

In this work, we first propose a novel beamforming technique for MIMO-OFDMA systems. By properly decomposing the {\em time-domain} channel response matrix into a product of subcarrier-dependent components and subcarrier-independent components, we can divide the beamforming task into the subcarrier-independent part implemented by the BS and the subcarrier-dependent part accomplished by each MT locally. Next, this novel beamforming technique is proposed to be incorporated into the design of opportunistic and beamforming schemes, which results in opportunistic and beamforming schemes that only require feedback of one beamforming matrix (BFM) index and $Q$ supportable throughputs for a MIMO-OFDMA system with $Q$ subcarriers. For comparison purposes, opportunistic and beamforming schemes employing the conventional SVD technique are also proposed by extending \cite{Chung03} to MIMO-OFDMA systems. While the SVD-based schemes can achieve the optimal throughput performance, they incur feedback overhead of $Q$ BFM indices and $Q$ supportable throughputs. Analytical and simulation results show that the proposed reduced-feedback schemes can asymptotically achieve the system throughput obtained by the SVD-based schemes.

\underline{Notation}: Vectors and matrices are denoted by boldface letters.  $\left\|\cdot\right\|$ represents the Euclidean norm of the enclosed vector and $\left|\cdot\right|$ denotes the amplitude of the enclosed complex-valued quantity. ${\bm I}_N$ is the $N\times N$ identity matrix. We use $E\left\{\cdot\right\}$, $\left(\cdot\right)^T$ and $\left(\cdot\right)^H$ for expectation, transposition and Hermitian transposition, respectively.

\section{Signal Model}\label{sec:smodel}
We consider a MIMO-OFDMA downlink system with $Q$ subcarriers and $K$ active MTs. The BS and each MT are equipped with $M$ and $N$ antennas with $M\leq N$, respectively. The propagation path between each pair of transmit and receive antennas is assumed to undergo independent slow frequency-selective fading and has a maximal channel length $L$. We denote by ${\bm h}_k^{m,n}(p)$ the $k$-th MT's channel impulse response between the $m$-th transmit antenna and the $n$-th receive antenna during the $p$-th OFDMA block. Assuming that the cyclic prefix (CP) is sufficient and synchronization has been achieved, the received signal by the $k$-th MT over its $q$-th subcarrier during the $p$-th OFDMA block can be written as
\begin{equation}\label{eq:rx}
{\bm y}_{k,q}(p)={\bm G}_{k,q}(p)\cdot{\bm x}_{k,q}(p)+{\bm w}_{k,q}(p),
\end{equation}
where ${\bm x}_{k,q}(p)$ is the pre-coded data vector and ${\bm w}_{k,q}(p)$ is complex circularly symmetric white Gaussian noise with zero-mean and unity variance. Furthermore, ${\bm G}_{k,q}(p)$ is the corresponding frequency-domain channel matrix computed as
\begin{equation}
{\bm G}_{k,q}(p)=\left[\begin{array}{llll}{\bm e}_q^T{\bm h}_k^{1,1}(p)&{\bm e}_q^T{\bm h}_k^{2,1}(p)&\cdots&{\bm e}_q^T{\bm h}_k^{M,1}(p)\\{\bm e}_q^T{\bm h}_k^{1,2}(p)&{\bm e}_q^T{\bm h}_k^{2,2}(p)&\cdots&{\bm e}_q^T{\bm h}_k^{M,2}(p)\\\vdots&\vdots&\ddots&\vdots\\{\bm e}_q^T{\bm h}_k^{1,N}(p)&{\bm e}_q^T{\bm h}_k^{2,N}(p)&\cdots&{\bm e}_q^T{\bm h}^{M,N}(p)\\\end{array}\right],
\end{equation}
with ${\bm e}_i$ being the vector containing the first $L$ elements of the $i$-th column of a $Q$-point discrete Fourier transform (DFT) matrix ${\bm W}$ whose entry is given by $\left[{\bm W}\right]_{\ell,u}=\exp \left(\frac{-j2\pi \ell u}{Q}\right)$ for $0\leq \ell, u \leq Q-1$.

In the following, we concentrate on the $k$-th MT over the $p$-th OFDMA block. For presentational clarity, we omit the MT and block indices, i.e. $k$ and $p$, in the sequel.

\section{Proposed Schemes}

\begin{figure*}[t]
\begin{center}
\includegraphics[scale=0.8]{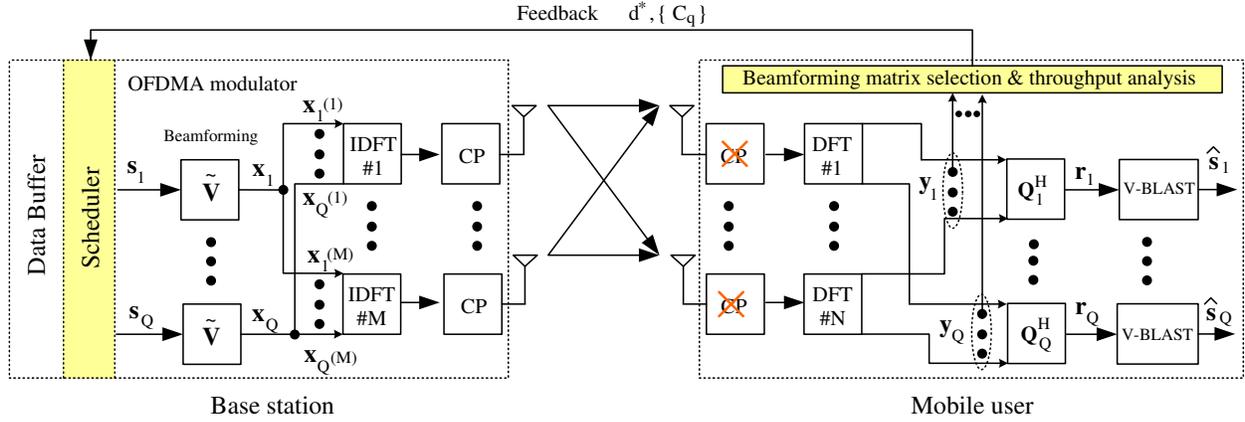}
\caption{Schematic diagram of the proposed per-subcarrier reduced-feedback opportunistic scheduling and beamforming scheme (PS-RF-OS).}\label{fig:system}
\end{center}
\end{figure*}

\subsection{Per-subcarrier reduced-feedback opportunistic and beamforming scheduling (PS-RF-OS)}
To begin with, we first rewrite ${\bm G}_q$ into the following form.
\begin{eqnarray}
{\bm G}_{q}&=&\left[{\bm I}_N\otimes{\bm e}_q^T\right]\left[\begin{array}{llll}{\bm h}^{1,1}&{\bm h}^{2,1}&\cdots&{\bm h}^{M,1}\\{\bm h}^{1,2}&{\bm h}^{2,2}&\cdots&{\bm h}^{M,2}\\\vdots&\vdots&\ddots&\vdots\\{\bm h}^{1,N}&{\bm h}^{2,N}&\cdots&{\bm h}^{M,N}\\\end{array}\right],\\
&=&\left[{\bm I}_N\otimes{\bm e}_q^T\right]{\bm H}, \label{eq:GIeH}
\end{eqnarray}
where $\otimes$ denotes the Kronecker product.

Next, ${\bm H}$ is singular-value decomposed as
\begin{equation}\label{eq:Hsvd}
{\bm H}={\bm U}\cdot{\bm \Sigma}\cdot{\bm V}^H,
\end{equation}
where ${\bm U}$ and ${\bm V}$ are unitary matrices of dimension $NL\times NL$ and $M\times M$, respectively. Furthermore, ${\bm \Sigma}$ is a diagonal matrix with $\min\left\{NL,M\right\}$ non-zero diagonal elements.

Substituting (\ref{eq:Hsvd}) into (\ref{eq:GIeH}), we have
\begin{eqnarray}
{\bm G}_q&=& \left[{\bm I}_N\otimes{\bm e}_q^T\right]\cdot{\bm U}\cdot{\bm \Sigma}\cdot{\bm V}^H\\
&=&{\bm Q}_q\cdot{\bm R}_q\cdot{\bm V}^H,\label{eq:GQRV}
\end{eqnarray}
where the last equality is obtained by the following QR decomposition :
\begin{equation}
\left[{\bm I}_N\otimes{\bm e}_q^T\right]\cdot{\bm U}\cdot{\bm \Sigma}={\bm Q}_q\cdot{\bm R}_q,
\end{equation}
with ${\bm Q}_q$ being a unitary matrix of dimension $M\times M$ and ${\bm R}_q$ an upper triangular matrix of dimension $M\times N$.

It is important to observe that ${\bm V}$ is independent of the subcarrier index $q$ and subsequently, can be employed as the BFM at the BS for all subcarriers. Furthermore, the beamforming for each subcarrier is characterized by ${\bm Q}_q$ which is known to the MT. Finally, in contrast to the conventional SVD-based channel decomposition where the channel matrix is decomposed as a product of two unitary matrices and a diagonal matrix, the diagonal matrix is replaced by an upper triangular matrix ${\bm R}_q$ in (\ref{eq:GQRV}). By exploiting these characteristics of (\ref{eq:GQRV}), we propose the following reduced-feedback opportunistic scheme for MIMO-OFDMA.

Assuming perfect channel estimation is achieved by each MT, we find a unitary BFM $\tilde{\bm V}$ satisfying ${\bm V}^H\tilde{\bm V}_{d}={\bm I}_M$ from a common codebook of $2^B$ entries shared by the BS and all MTs. We define
\begin{equation}\label{eq:gamma}
\gamma_{i,j}\defequal\left|\left[{\bm V}^H\tilde{\bm V}_d\right]_{i,j}\right|^2.
\end{equation}
It is easy to show that $0\leq \gamma_{m,m}\leq 1$ for $m=1,2,\cdots,M$. The following criterion is employed to select $\tilde{\bm V}_d$.
\begin{equation}\label{eq:dstar}
d^*=\argmax_{d\in\left[1,2^B\right]}\sum_{m=1}^M{\frac{\gamma_{m,m}}{\sum_{m\neq j}\gamma_{m,j}}}.
\end{equation}

Obviously, it is more likely to achieve ${\bm V}^H\tilde{\bm V}_{d^*}={\bm I}_M$ if the codebook contains more candidate BFMs at the cost of more feedback bits.

The integer $d^*$ is then returned to the BS and $\tilde{\bm V}_{d^*}$ is employed to pre-coded the data symbols as
\begin{equation}
{\bm x}_q=\tilde{\bm V}_{d^*}\cdot{\bm s}_q,
\end{equation}
where ${\bm s}_q$ is the data vector of length $M$ and $E\left\{\left\|{\bm s}_q\right\|^2\right\}=\rho M$.

Assuming ${\bm V}^H\tilde{\bm V}_{d^*}={\bm I}_M$ and recalling (\ref{eq:GQRV}), we pre-multiply ${\bm Q}_q^H$ on both sides of (\ref{eq:rx}) and have
\begin{eqnarray}
{\bm r}_q&=&{\bm Q}_q^H\cdot{\bm y},\\
&=&{\bm R}_q\cdot{\bm s}_q+{\bm w}'_q,\label{eq:rQH}
\end{eqnarray}
where ${\bm w}'_q={\bm Q}_q^H{\bm w}_q$.

Since ${\bm R}_q$ is an upper triangular matrix, the vertical Bell Laboratories Layered Space-Time (V-BLAST) receiver \cite{Wolniansky98} can be employed to detect ${\bm s}_q(m)$ for $m=M,M-1,\cdots,1$ sequentially, where ${\bm s}_q(m)$ is the $m$-th entry of ${\bm s}_q$. It has been shown in \cite{Jiang05} that the adverse effect of error propagation in V-BLAST receivers is negligible for the high SNR region. As a result, (\ref{eq:rQH}) can be rewritten as $M$ equivalent parallel channels:
\begin{equation}
\left[{\bm r}_q\right]_m=\left[{\bm R}_q\right]_{m,m}\left[{\bm s}_q\right]_m+\left[{\bm w}'_q\right]_m,\quad m=1,2,\cdots,M,
\end{equation}
where $\left[{\bm R}_q\right]_{m,m}$ is the $m$-th diagonal element of ${\bm R}_q$. Thus, the supportable data throughput on the $q$-th subcarrier is given by
\begin{equation}\label{eq:tildeCq}
\tilde{C}_q=\sum_{m=1}^M\log_2\left(1+\rho\left|\left[{\bm R}_q\right]_{m,m}\right|^2\right).
\end{equation}

However, only ${\bm V}^H\tilde{\bm V}\approx{\bm I}_M$ holds for practical systems with a finite codebook. Thus, the actual supportable throughput can be approximated as
\begin{equation}\label{eq:Cq}
C_q\approx\sum_{m=1}^M\log_2\left(1+\rho\left|\left[{\bm R}_q\right]_{m,m}\right|^2\gamma_{m,m}\right).
\end{equation}

Finally, each MT returns to the BS one integer BFM index $d^*$ as well as $Q$ real-valued supportable throughputs $\left\{C_q; q=1,2,\cdots,Q\right\}$. After receiving information from all MTs, the BS assigns each subcarrier to the MT with the highest supportable data throughput. In the sequel, this scheme is referred to as the per-subcarrier reduced-feedback opportunistic scheduling and beamforming scheme (PS-RF-OS). PS-RF-OS is depicted in Fig. \ref{fig:system} and summarized as follows.

\begin{enumerate}
\item In the beginning of a time slot, each MT estimates its time-domain channel response matrix $\bm H$ by exploiting a pilot signal broadcast from the BS;
\item Each MT computes ${\bm Q}_q$, ${\bm R}_q$ and ${\bm V}$ according to (\ref{eq:GQRV});
\item The best BFM $\tilde{\bm V}_{d^*}$ is selected from the common codebook based on the selection criterion (\ref{eq:dstar});
\item Finally, each MT feeds $d^*$ and $\left\{C_q\right\}$ back to the BS;
\item Taking fairness in account, the BS assigns subcarriers to each MT based on $\left\{C_q\right\}$;
\item The BS transmits data pre-coded with $\tilde{\bm V}_{d^*}$ throughout the current slot;
\item The above procedures are repeated in the next slot.
\end{enumerate}

\subsection{Per-cluster reduced-feedback opportunistic and beamforming scheduling (PC-RF-OS)}

Inspired by \cite{Svedman04}, a further feedback reduction can be achieved by dividing the $Q$ subcarriers into $G$ exclusive clusters composed of $U$ adjacent subcarriers with $Q=G\times U$. Assuming that the channel conditions are approximately constant over the $U$ subcarriers in the same cluster, the following average supportable data throughput of each cluster is fed back to the BS:
\begin{equation}
\bar{C}_g=\frac{1}{U}\sum_{q\in {\cal I}_g}{C}_q,
\end{equation}
where ${\cal I}_g=\left\{i_{g,1},i_{g,2},\cdots,i_{g,U}\right\}$ is the subcarrier index set of the $g$-th cluster.

As a result, only one integer-valued $d^*$ and $G$ real-valued supportable throughputs are required to be fed back to the BS. In the sequel, this scheme is referred to as the per-cluster reduced-feedback opportunistic scheduling and beamforming scheme (PC-RF-OS).

\subsection{Per-subcarrier eigen-beamforming opportunistic scheduling (PS-EB-OS)}
For comparison purposes, two eigen-beamforming scheduling schemes are proposed in the following. These two schemes can be considered as a MIMO-OFDMA extension of the SC-MIMO scheme proposed in \cite{Chung03} and require the conventional amount of feedback. For each subcarrier, its channel matrix ${\bm G}_{q}$ is decomposed as
\begin{equation}\label{eq:Gqsvd}
{\bm G}_{q}={\bm U}_q\cdot{\bm \Sigma}_q\cdot{\bm V}_q^H,
\end{equation}
where ${\bm U}_q$ and ${\bm V}_q$ are unitary matrices of dimension $N\times N$ and $M\times M$, respectively. Furthermore, ${\bm \Sigma}_q$ is a diagonal matrix whose first $M$ diagonal elements are the singular values of ${\bm G}_{q}$, $\lambda_{q,i}$ for $i=1,2,\cdots,M$. In contrast to (\ref{eq:GQRV}), ${\bm V}_q$ is {\em dependent} on the subcarrier index $q$.

Thus, the transmitted signal can be pre-coded with a BFM $\tilde{\bm V}_q$ chosen from the common codebook in an approach similar to (\ref{eq:dstar}). Assuming ${\bm V}^H\tilde{\bm V}_q={\bm I}$, we pre-multiply the received signal in (\ref{eq:rx}) with ${\bm U}_q^H$ and obtain
\begin{equation}
{\bm r}'_q={\bm U}_q^H{\bm y}={\bm \Sigma}_q{\bm s}_q+{\bm w}''_q,
\end{equation}
where ${\bm w}''_q={\bm U}_q^H{\bm w}_q$.

Finally, we can have the resulting supportable data throughput as
\begin{equation}\label{eq:tildeRq}
\tilde{T}_q=\sum_{m=1}^M\log_2\left(1+\rho\left|\lambda_{q,m}\right|^2\right).
\end{equation}

Since ${\bm V}_q^H\tilde{\bm V}_q\approx{\bm I}_M$ for a finite codebook, the actual supportable throughput can be approximated by
\begin{equation}\label{eq:Rq}
T_q\approx\sum_{m=1}^M\log_2\left(1+\rho\left|\lambda_{q,m}\right|^2\gamma'_{m,m}\right),
\end{equation}
where $\gamma'_{m,m}$ is evaluated as shown in (\ref{eq:gamma}) with $\bm V$ being replaced by $\bm V_q$.

It is clear that this approach requires each MT to feed back $Q$ BFM indices as well as $Q$ real-valued supportable data throughputs, which amounts to approximately twice the feedback amount of PS-RF-OS. In the sequel, this scheme is referred to as the per-subcarrier eigen-beamforming opportunistic scheduling and beamforming scheme (PS-EB-OS).

\subsection{Per-cluster eigen-beamforming opportunistic scheduling (PC-EB-OS)}
If subcarriers are grouped into clusters, then feedback can be reduced to only one BFM index and the average supportable throughput for each cluster. This results in a feedback reduction by a factor of $U$. In the sequel, this scheme is referred to as the per-cluster eigen-beamforming opportunistic scheduling and beamforming scheme (PC-EB-OS).

\subsection{Remarks}
The following comparison about the four proposed schemes is of interest.

\noindent 1.) PS-EB-OS/PC-EB-OS requires twice the amount of feedback of PS-RF-OS/PC-RF-OS, respectively;

\noindent 2.) In PC-EB-OS, each MT feeds one BFM index per cluster back to the BS. Consequently, the chosen BFM may be a poor approximation of the true eigen-BFM for some subcarriers in the cluster, which incurs severe performance degradation as the cluster size increases. On the contrary, PC-RF-OS entails only marginal performance degradation due to clustering since the feedback reduction involves no BFM information loss.

\noindent 3.) It is fair to say that the substantial feedback reduction of PS-RF-OS/PC-RF-OS is achieved at the cost of higher computational complexity at the MTs compared to PS-EB-OS/PC-EB-OS since the V-BLAST receivers have to be employed to detect the data symbols.

\section{Asymptotic Performance Analysis}\label{sec:analysis}
In this section, we compare the asymptotic data throughput of PS-RF-OS in (\ref{eq:tildeCq}) and that of PS-EB-OS in (\ref{eq:tildeRq}). Observing that (\ref{eq:GQRV}) and (\ref{eq:Gqsvd}) stand for two different decomposition expressions of the same matrix ${\bm G}_{q}$, we have
\begin{equation}
{\bm Q}_q\cdot{\bm R}_q\cdot{\bm V}^H={\bm U}_q\cdot{\bm \Sigma}_q\cdot{\bm V}_q^H.
\end{equation}

On taking the determinant of both sides and recalling that ${\bm Q}_q$, ${\bm V}$, ${\bm U}_q$ and ${\bm V}_q$ are all unitary matrices, we can easily obtain
\begin{equation}
\det\left({\bm R}_q\right)=\det\left({\bm \Sigma}_q\right),
\end{equation}
or equivalently,
\begin{equation}
\prod_{m=1}^M \left|\left[{\bm R}_q\right]_{m,m}\right|=\prod_{m=1}^M \left|\lambda_{q,m}\right|.
\end{equation}

Assuming $\rho\left|\lambda_{q,m}\right|\gg 1$ and $\rho\left|\left[{\bm R}_q\right]_{m,m}\right|\gg 1$, we can obtain the following asymptotic relationship between the two data throughputs as $\rho$ increases to infinity.
\begin{equation}
\frac{\tilde{C}_q}{\tilde{T}_q}\approx\frac{M\log_2\left(\rho\right)+2\log_2\left(\displaystyle\prod_{m=1}^M \left|\left[{\bm R}_q\right]_{m,m}\right|\right)}{M\log_2\left(\rho\right)+2\log_2\left(\displaystyle\prod_{m=1}^M \left|{\lambda}_{q,m}\right|\right)}=1.
\end{equation}
Thus, the data throughput of PS-RF-OS approaches to that of PS-EB-OS asymptotically as $\rho$ increases.

\section{Simulation Results}
Computer simulation are conducted in this section to assess the performance of the four proposed scheduling and beamforming schemes. The simulated OFDMA system has $Q=128$ subcarriers and $M=N=2$ antennas. The channel response of each MT is generated according to the HIPERLAN/2 channel model with eight paths ($L=8$). In particular, the channel coefficients are modeled as independent and complex-valued Gaussian random variables with zero-mean and an
exponential power delay profile
\begin{equation}\label{eq:simchn}
E\left\{\left|h_k(l)\right|^2\right\}=\lambda\cdot
\exp\left\{-l\right\},\quad{l=0,1,\cdots,7.}
\end{equation}
where the constant $\lambda$ is such chosen that $E\left\{\left|\bm h_k\right|^2 \right\}=1$. To prevent inter-block interference (IBI), a CP of length $N_g=8$ is employed. Notice that we have implicitly assumed a homogenous network in (\ref{eq:simchn}). As a result, allocating the subcarrier/cluster to the MT with the highest throughput is a fair scheduling strategy. Unless otherwise specified, we set $SNR=10$ dB, $K=10$, $B=8$ for all schemes and $G=8$ for PC-RF-OS and PC-EB-OS.

\begin{figure}[htp]
\begin{center}
\includegraphics[scale=0.43]{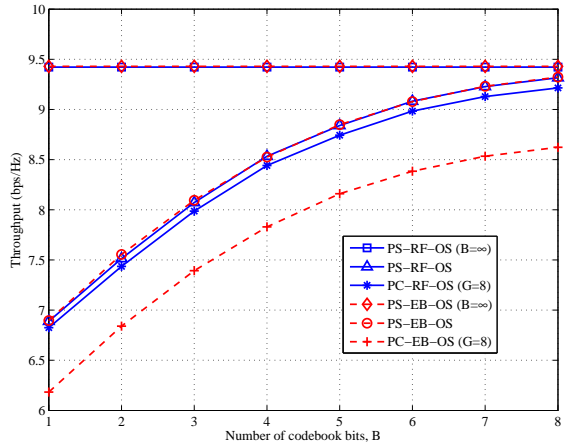}
\caption{Average throughput as a function of the number of bits in the codebook, $B$.}\label{fig:befffig}
\end{center}
\end{figure}

\subsection*{Case 1: Impact of the codebook size}
In this experiment, we investigate the impact of the codebook size on the system throughput. Figure \ref{fig:befffig} shows the average throughput as a function of $B$. The curves labeled ``$B=\infty$" serve as the performance upper bound with ${\bm V}^H\tilde{\bm V}_{d^*}={\bm I}_M$ or ${\bm V}_q^H\tilde{\bm V}_{d_q^*}={\bm I}_M$. Figure \ref{fig:befffig} indicates that the throughput of all schemes improves as the codebook size increases. In particular, Fig. \ref{fig:befffig} suggests that a codebook with $B=8$ bits is sufficiently large to allow PS-RF-OS, PS-EB-OS and PC-RF-OS to achieve throughput performance approaching that achieved with $B=\infty$. In other words, $B=8$ is sufficient to achieve ${\bm V}^H\tilde{\bm V}_{d^*}={\bm I}_M$ or ${\bm V}_q^H\tilde{\bm V}_{d_q^*}={\bm I}_M$.

\subsection*{Case 2: Throughput comparison}
\begin{figure}[htp]
\begin{center}
\includegraphics[scale=0.43]{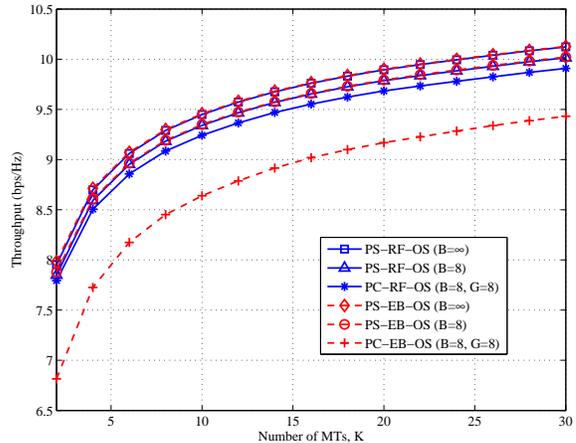}
\caption{Average throughput as a function of the number of MTs, $K$.}\label{fig:scalingfig}
\end{center}
\end{figure}

In the second experiment, we compare the performance of the four scheduling schemes in terms of throughput as a function of $K$, the number of MTs. It is clear from Fig. \ref{fig:scalingfig} that the throughput performance of PS-RF-OS and PS-EB-OS with ``$B=\infty$" is identical, which confirms our asymptotic analysis in Sec. \ref{sec:analysis}. Furthermore, the performance difference among PS-RF-OS, PS-EB-OS and PC-RF-OS with $B=8$ is marginal whereas the throughput of PC-EB-OS suffers from the reduced feedback.

\subsection*{Case 3: Impact of the cluster size}

\begin{figure}[htp]
\begin{center}
\includegraphics[scale=0.43]{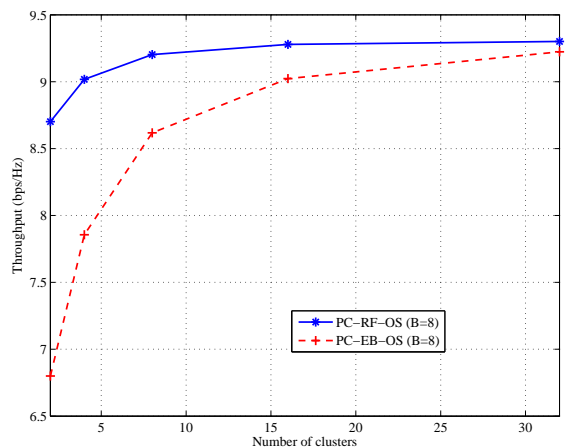}
\caption{Average throughput as a function of the number of clusters, $G$.}\label{fig:geffig}
\end{center}
\end{figure}

In this last experiment, we study the impact of the cluster size on the system throughput of PC-RF-OS and PC-EB-OS. Figure \ref{fig:geffig} shows that PC-RF-OS entails less performance degradation due to adopting a larger cluster size, i.e. a smaller $G$, compared to PC-EB-OS. This is because PC-RF-OS only reduces the feedback information on the supportable throughput of each subcarrier whereas PC-EB-OS further suffers from the information loss about the optimal BFM.

\section{Conclusion}
Two opportunistic scheduling and beamforming schemes, namely PS-RF-OS and PC-RF-OS, have been proposed for MIMO-OFDMA downlink systems by exploiting a novel beamforming technique where beamforming is performed jointly by both the BS and MTs. The resulting PS-RF-OS requires feedback of only one BFM index and $Q$ supportable throughputs for a MIMO-OFDMA system with $Q$ subcarriers. A further feedback reduction is achieved by PC-RF-OS through grouping adjacent subcarriers into clusters and returning only cluster information from each MT. Since PC-RF-OS does not incur feedback loss regarding the BFM information, it only entails marginal performance degradation with respect to PS-RF-OS. For comparison purposes, two SVD-based opportunistic scheduling and beamforming schemes, namely PS-EB-OS and PC-EB-OS, have been developed. The superior performance of PS-EB-OS is obtained at the cost of twice much amount of feedback compared to PS-RF-OS and PC-RF-OS whereas the performance of PC-EB-OS decreases rapidly as the cluster size increases. It has been demonstrated through asymptotic analysis and computer simulation that PS-RF-OS and PC-RF-OS can achieve throughput performance comparable to PS-EB-OS with substantially reduced feedback.

\bibliographystyle{IEEEtranS}
\bibliography{Bib}

\begin{thebibliography}{1}
\providecommand{\url}[1]{#1}
\csname url@samestyle\endcsname
\providecommand{\newblock}{\relax}
\providecommand{\bibinfo}[2]{#2}
\providecommand{\BIBentrySTDinterwordspacing}{\spaceskip=0pt\relax}
\providecommand{\BIBentryALTinterwordstretchfactor}{4}
\providecommand{\BIBentryALTinterwordspacing}{\spaceskip=\fontdimen2\font plus
\BIBentryALTinterwordstretchfactor\fontdimen3\font minus
  \fontdimen4\font\relax}
\providecommand{\BIBforeignlanguage}[2]{{%
\expandafter\ifx\csname l@#1\endcsname\relax
\typeout{** WARNING: IEEEtranS.bst: No hyphenation pattern has been}%
\typeout{** loaded for the language `#1'. Using the pattern for}%
\typeout{** the default language instead.}%
\else
\language=\csname l@#1\endcsname
\fi
#2}}
\providecommand{\BIBdecl}{\relax}
\BIBdecl

\bibitem{Chung03}
J.~Chung, C.~Hwang, K.~Kim, and Y.~K. Kim, ``A random beamforming technique in
  {MIMO} systems exploiting multiuser diversity,'' \emph{IEEE Journal Select.
  Areas Commun.}, vol.~21, no.~5, pp. 848--855, Jun. 2003.

\bibitem{Wolniansky98}
G.~D. Golden, C.~J. Foschini, and P.~W. Wolniansky, ``Detection algorithm and
  initial laboratory results using {V-BLAST} space-time communication
  architecture,'' \emph{Electronics Letters}, vol.~35, pp. 14--16, Jan. 1999.

\bibitem{Kim05}
I.~Kim, S.~Hong, S.~S. Ghassemzadeh, and V.~Tarokh, ``Opportunistic beamforming
  based on multiple weighting vectors,'' \emph{IEEE Trans. Wireless Commun.},
  vol.~4, no.~6, pp. 2683--2687, November 2005.

\bibitem{Svedman04}
P.~Svedman, S.~K. Wilson, L.~J. Cimini, Jr., and B.~Ottersten, ``A simplified
  opportunistic feedback and scheduling scheme for {OFDM},'' \emph{Proceedings
  {IEEE} Vehicular Technology Conference, Spring}, vol.~4, pp. 1878--1882, May
  2004.

\bibitem{Tse02}
P.~Viswanath, D.~N.~C. Tse, and R.~Laroia, ``Opportunistic beamforming using
  dumb antennas,'' \emph{IEEE Trans. Info. Theory}, vol.~48, no.~6, pp.
  1277--1294, Jun. 2002.

\bibitem{Wong99}
C.~Y. Wong, R.~S. Cheng, K.~B. Lataief, and R.~D. Murch, ``Multiuser {OFDM}
  with adaptive subcarrier, bit, and power allocation,'' \emph{IEEE Journal
  Sel. Area in Comm.}, vol.~17, pp. 1747--1758, Oct. 1999.

\end{thebibliography}
\end{document}